# Role of electron-electron interactions in the charge dynamics of rare-earth-doped $CaFe_2As_2$

Zhen Xing, T. J. Huffman, Peng Xu, A. J. Hollingshad, D. J. Brooker, N. E. Penthorn, M. M. Qazilbash*

*Department of Physics, College of William and Mary, Williamsburg, Virginia 23187-8795, USA*

S. R. Saha, T. Drye, C. Roncaioli, J. Paglione

*Center for Nanophysics and Advanced Materials, Department of Physics,*
*University of Maryland, College Park, Maryland 20742, USA*

We have investigated the charge dynamics and the nature of many-body interactions in La- and Pr- doped $CaFe_2As_2$. From the infrared part of the optical conductivity, we discover that the scattering rate of mobile carriers above 200 K exhibits saturation at the Mott-Ioffe-Regel limit of metallic transport. However, the dc resistivity continues to increase with temperature above 200 K due to the loss of Drude spectral weight. The loss of Drude spectral weight with increasing temperature is seen in a wide temperature range in the uncollapsed tetragonal phase, and this spectral weight is recovered at energy scales about one order of magnitude larger than the Fermi energy scale in these semimetals. The phenomena noted above have been observed previously in other correlated metals in which the dominant interactions are electronic in origin. Further evidence of significant electron-electron interactions is obtained from the presence of quadratic temperature and frequency-dependent terms in the scattering rate at low temperatures and frequencies in the uncollapsed tetragonal structures of La-doped and Pr-doped $CaFe_2As_2$. For temperatures below the structure collapse transition in Pr-doped $CaFe_2As_2$ at ~70 K, the scattering rate decreases due to weakening of electronic correlations and the Drude spectral weight decreases due to modification of the low energy electronic structure.

## I. INTRODUCTION

The investigation of magnetic, structural, transport, and superconducting properties of pure and doped crystals of the 122 family of iron arsenides $AFe_2As_2$ ($A$ = Ba, Ca, Sr) has played a pivotal role in furthering our understanding of the fascinating many-body interactions and phase transitions observed in the iron pnictides and iron chalcogenides [1–3]. The parent compounds in the 122 family go through a tetragonal-to-orthorhombic structural transition coupled with antiferromagnetic spin-density-wave order at low temperatures [1–3]. In $CaFe_2As_2$, pressure-induced superconductivity emerges only under nonhydrostatic experimental conditions [4–6]. Under hydrostatic pressure, instead of superconductivity, a so-called collapsed tetragonal (CT) phase occurs resulting in a dramatic *c*-axis reduction (about 10%) without breaking symmetry [7,8]. $CaFe_2As_2$ is much more sensitive to stress anisotropy compared to $BaFe_2As_2$ and $SrFe_2As_2$ which also show pressure-induced superconductivity and CT phase but at much higher pressure [9–12].

The CT phase, which is driven by interlayer As-As separation [13,14], can also be stabilized by chemical substitution in $CaFe_2As_2$ at ambient pressure. The antiferromagnetism in $CaFe_2As_2$ is suppressed by appropriate doping, for example, by substituting rare-earth Pr and Nd on the Ca site [14], Rh on the iron site [15], or phosphorus on the As site [16], leading to the emergence of the CT phase. Depending upon the trivalent rare-earth-ion substitution in the system [14], $CaFe_2As_2$ can either maintain the uncollapsed tetragonal (UT) structure with La substituent, or undergo a phase transition at low temperature from the UT structure to the CT structure with Nd or Pr substituents. Hence, rare-earth doped $CaFe_2As_2$ crystals provide us the chance to study (in a controlled manner) the UT and CT phases at ambient pressure [17–21]. The rare-earth substituents are believed to dope electrons into the system in addition to varying the chemical pressure due to their different ionic radii compared to the calcium ion. In the CT phase, Fe local moments are quenched [7,18], spin fluctuations are suppressed [22], and electron correlations are believed to be reduced [23]. Angle-resolved photoemission spectroscopy (ARPES) results show that there is reconstruction of the Fermi surface in the CT structure in strained crystals of $CaFe_2As_2$, including the complete disappearance of the hole pocket at the zone center (Γ point) [24,25], consistent with theoretical expectation [13,26]. However, very recent ARPES experiments on Pr-doped $CaFe_2As_2$ show that across the CT phase transition, the hole pocket at Γ point does not disappear completely [27], which is different from the CT phase in the parent compound under internal strain [24,25]. The added diversity in the rare-earth doped $CaFe_2As_2$ system provides us the opportunity to study with optical spectroscopy the electronic structure and the nature of many-body interactions. Unlike previous infrared work [28], we investigate both $Ca_{0.8}La_{0.2}Fe_2As_2$ and $Ca_{0.85}Pr_{0.15}Fe_2As_2$ crystals to compare the properties of the UT phase of the former material with the UT and CT phases of the latter material.

In this paper, the frequency- and temperature-

---

dependent *ab*-plane optical constants of $Ca_{0.8}La_{0.2}Fe_2As_2$ and $Ca_{0.85}Pr_{0.15}Fe_2As_2$ crystals are obtained through optical spectroscopy. An interesting finding is that the scattering rate saturates above ~ 200 K in the UT structure in La-doped and Pr-doped $CaFe_2As_2$. However, the resistivity continues to increase above 200 K, which we find to be a consequence of the loss of mobile carriers. The loss of Drude spectral weight of mobile carriers with increasing temperature is seen in a wide temperature range in the uncollapsed tetragonal phase, and this spectral weight is recovered about 0.5 eV, much larger than the Fermi energy scale in these semimetals. The scattering rate in La-doped $CaFe_2As_2$ between 5 K and 150 K is dominated by a quadratic temperature-dependent term ascribed to significant electron-electron interactions. The frequency dependence of the scattering rate obtained from the extended Drude analysis is in accord with its temperature dependence. We document the impact of the structure collapse transition on the infrared properties of the Pr doped system, and also compare these properties with those of the UT phase of La doped $CaFe_2As_2$. We find that the plasma frequency and scattering rate of free carriers decrease across the CT phase transition. Optical interband transitions are also affected by electronic structure reconstruction across the CT phase transition.

## II. SAMPLES AND EXPERIMENTS

Single crystals of $Ca_{0.8}La_{0.2}Fe_2As_2$ and $Ca_{0.85}Pr_{0.15}Fe_2As_2$ were grown using the FeAs self-flux method [14]. At these rare-earth doping levels, the spin density wave transition is suppressed. The temperature-dependent resistivity data for $Ca_{0.8}La_{0.2}Fe_2As_2$ shows metallic behavior with no signs of a magnetic or structural phase transition. In the Pr-doped samples, the CT phase occurs below 70 K with a subtle kink in the resistivity curve and a dramatic change in the Hall coefficient [14]. The size of $Ca_{0.8}La_{0.2}Fe_2As_2$ crystals is as large as $10\times10\times2$ mm$^3$, and the size of $Ca_{0.85}Pr_{0.15}Fe_2As_2$ crystals is as large as $5\times5\times1$ mm$^3$. It is easy to obtain relatively flat and shiny *ab*-plane surfaces by cleaving.

Near-normal incidence reflectance measurements on the *ab*-plane surfaces were performed with the Bruker Vertex 80v Fourier transform infrared (FTIR) spectrometer in the frequency range 60 cm$^{-1}$ – 8000 cm$^{-1}$ and temperature range 5 K – 300 K (Appendix A). An *in situ* gold evaporation method similar to that described in Ref. [29] was used to obtain absolute reflectance. Ellipsometry measurements on the *ab*-plane surfaces were performed with a Woollam variable-angle spectroscopic ellipsometer (VASE) in the frequency range 4800 cm$^{-1}$ – 40 000 cm$^{-1}$ and temperature range 5 K – 300 K (Appendix A). In this frequency range, the complex optical conductivity was obtained directly from the measured ellipsometric coefficients. The infrared conductivity at lower frequencies is obtained by Kramers-Kronig (KK) transformation on reflectance constrained by ellipsometry results [30]. Both Hagen-Rubens and Drude extrapolations [31] constrained by dc conductivity of the crystals were employed at very low frequencies in order to perform KK transformations (Appendix A). The optical constants obtained in the frequency range of interest are hardly affected by the choice of the very low frequency extrapolation function.

## III. RESULTS AND DISCUSSION

### A. Optical conductivity and spectral weight

The real part ($\sigma_1$) of the optical conductivity is shown in Fig. 1. The $Ca_{0.8}La_{0.2}Fe_2As_2$ crystal shows metallic behavior at low temperatures with a clear Drude-like feature at low frequencies. However, at higher temperatures, there is a nonmonotonic frequency dependence that appears to depart from Drude-like conductivity. For the Pr-doped $CaFe_2As_2$ crystal, spectra have been measured between 300 K and 100 K in the UT phase, and at 40 K and 5 K in the CT phase.

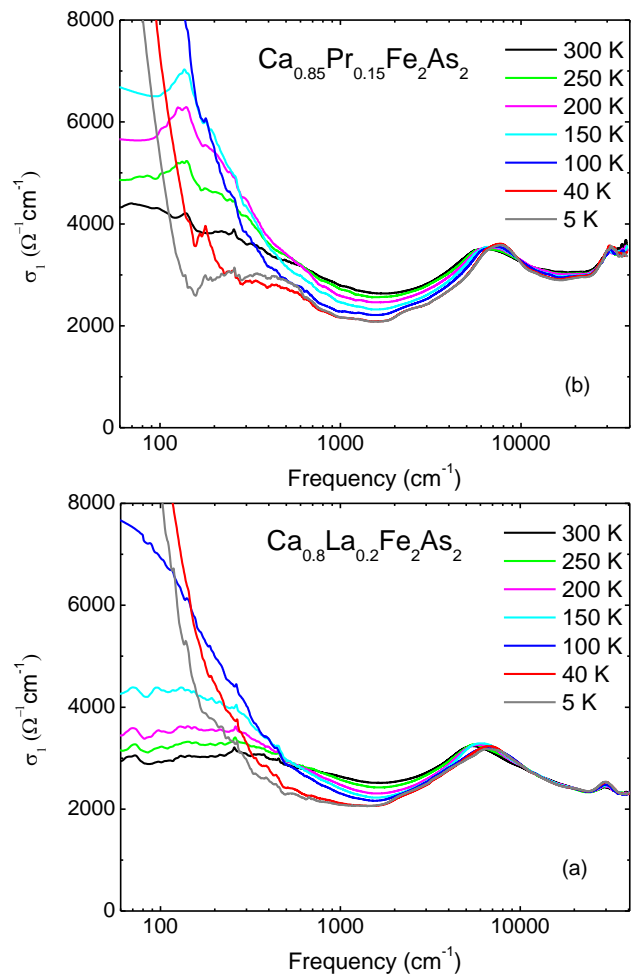

FIG. 1. The real part of the *ab*-plane optical conductivity $\sigma_1$ is plotted as a function of frequency at different temperatures for (a) $Ca_{0.8}La_{0.2}Fe_2As_2$ and (b) $Ca_{0.85}Pr_{0.15}Fe_2As_2$.

The optical conductivity in both phases is consistent with metallic behavior. The occurrence of the CT phase transition is apparent in the shift of the infrared-active, Fe-As phonon center frequency (Appendix B).

We calculate the spectral weight (SW) as a function of frequency via the integral of $\sigma_1$ for both materials:

$$\text{SW}(\omega) = \int_0^\omega \sigma_1(\omega')d\omega'. \quad (1)$$

This integral is calculated for optical conductivities at different temperatures. For conducting materials at low frequencies, the spectral weight is proportional to square of the plasma frequency, and hence the number of charge carriers in the material [31]. If we assume the charge carriers have masses equal to the free electron mass, we may rewrite the spectral weight in terms of an effective number of carriers $N_{\text{eff}}$ per formula unit in a primitive cell volume $V_0$:

$$N_{\text{eff}}(\omega) = \frac{2m_e V_0}{\pi e^2} \int_0^\omega \sigma_1(\omega')d\omega'. \quad (2)$$

The effective number of carriers are shown in Fig. 2. It is clear that at lower frequencies, the spectral weight decreases with increasing temperature. Phase space restrictions for the holelike bands in these semimetals due to the Pauli exclusion principle may contribute to spectral weight redistribution on the order of the Fermi energy (~ 0.05 eV or 400 cm$^{-1}$). However, we note that the total spectral weight is conserved at about 4000 cm$^{-1}$ (~ 0.5 eV) for the data in the UT phase. This spectral weight recovery energy scale is about one order of magnitude larger than the Fermi energy scales (~ 0.02 eV – 0.07 eV) of the electron and hole carriers in the rare-earth-doped CaFe$_2$As$_2$. Interactions between charge carriers redistribute the spectral weight to energies much higher than the Fermi energies. We also note that the energy scale over which the spectral weight is recovered is not too different from that seen in the cuprates (~ 2 eV) [32]. In our work, the Fermi energy is defined from the Fermi level to the bottom of the electronlike bands (for electron pockets) or the top of the holelike bands (for hole pockets). In other words, the Fermi energy is either the occupied bandwidth of the electronlike bands or the unoccupied bandwidth of the holelike bands.

We fit the complex conductivity with the Drude-Lorentz model [31]:

$$\sigma(\omega) = \frac{\omega_p^2}{4\pi}\frac{1}{1/\tau - i\omega} + \sum_j \frac{\Omega_j^2}{4\pi}\frac{\omega}{i(\omega_j^2 - \omega^2) + \omega/\tau_j}, \quad (3)$$

where the first term is the Drude component which represents free-carrier response, and latter terms are Lorentz components, which represent the response associated with localized charges and/or optical interband transitions. In the UT structures of both samples, we find that only one Drude term and one overdamped, midinfrared Lorentz oscillator is sufficient for a very good low-frequency fit (Appendix C). This fitting procedure for the infrared conductivity has been used previously in the literature [33,34]. Due to the multiband nature of iron-based materials [2,3], it is usually more difficult to interpret the infrared conductivity. Other researchers have fit their data with two Drude terms in which one is narrow and the other is very broad [35–37]. However, the two-Drude model does not provide satisfactory fits to our infrared data at higher temperatures as we show in Appendix C. Moreover, the scattering rate parameter of the broad Drude appears to be unphysical [34] because it is several times the value of the Fermi energies of the electron and hole carriers.

### B. Free carrier response

We first focus on the Drude component, which represents the free-carrier response. The square of the plasma frequency $\omega_p^2$ and scattering rate $1/(2\pi c\tau)$ normalized to the respective values at 300 K are shown in Fig. 3(a) and 3(b) as a function of temperature. A discontinuity in the magnitude of the plasma frequency occurs below the CT transition temperature of Pr-doped $CaFe_2As_2$, which implies a discontinuous reduction of carrier density. This is consistent with the ARPES results on $Ca_{0.85}Pr_{0.15}Fe_2As_2$, which show a significant reduction of a large hole pocket and the disappearance of the small hole pocket after structure collapse [27]. Recently, another structure collapse material $CaFe_2(As_{0.935}P_{0.065})_2$ has been studied, as described in Ref. [38]. In $CaFe_2(As_{0.935}P_{0.065})_2$, a noticeable suppression of reflectance occurs between 1000 cm$^{-1}$ and 3500 cm$^{-1}$, resulting in a deeper valley about 1500 cm$^{-1}$ in $\sigma_1$ in the CT phase. This behavior is nominally different from that observed in $Ca_{0.85}Pr_{0.15}Fe_2As_2$ in our work, probably due to differences in details of the electronic structure. However, similar to $Ca_{0.85}Pr_{0.15}Fe_2As_2$, in $CaFe_2(As_{0.935}P_{0.065})_2$ the (total) plasma frequency of Drude contribution decreases across the CT phase transition.

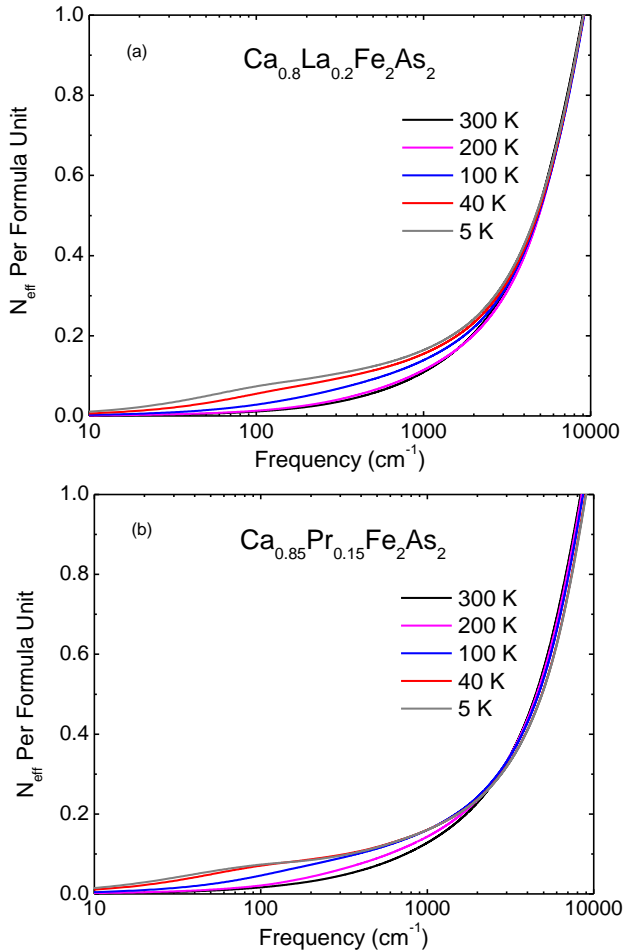

FIG. 2. Effective number of carriers $N_{\text{eff}}$ at different temperatures for (a) Ca$_{0.8}$La$_{0.2}$Fe$_2$As$_2$ and (b) Ca$_{0.85}$Pr$_{0.15}$Fe$_2$As$_2$.

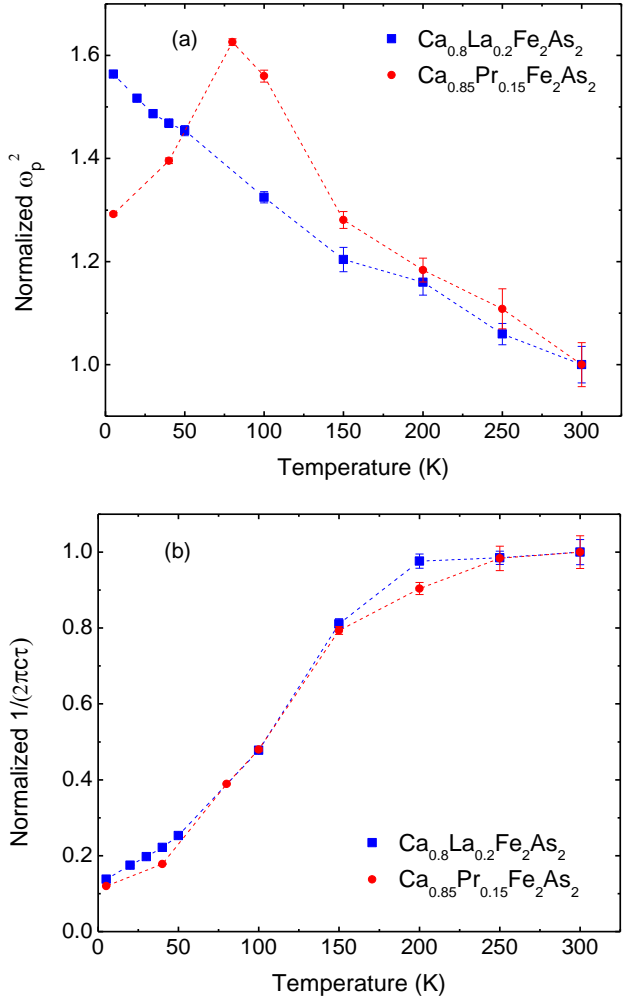

FIG. 3. Temperature dependence of the parameters of the Drude term (a) $\omega_p^2$ and (b) $1/(2\pi c\tau)$ normalized to the respective values at 300 K for $Ca_{0.8}La_{0.2}Fe_2As_2$ (blue squares) and $Ca_{0.85}Pr_{0.15}Fe_2As_2$ (red circles). Dashed lines are guides to the eye.

Remarkably, scattering rate of both La- and Pr-doped $CaFe_2As_2$ shows saturation above 200 K, clearly indicating the attainment of the Mott-Ioffe-Regel limit of metallic transport. However, the resistivity continues to increase above 200 K as shown in Ref. [14]. We find this to be a consequence of the decrease in number density of mobile carriers and is directly seen in the decrease of the Drude spectral weight (square of the plasma frequency) in Fig. 3(a). The decrease of the Drude spectral weight with increasing temperature is consistent with the model independent analysis shown in Fig. 2(a), and 2(b) and discussed in the preceding section.

Another criterion for the Mott-Ioffe-Regel limit is that the quantity $k_F l$ is of order unity. This quantity can be estimated by the resistivity formula for two-dimensional systems [39]:

$$\rho_{2D} = \frac{2\pi\hbar c_0}{e^2 k_F l}\frac{1}{M}. \qquad (4)$$

We find that $k_F l \sim 1$ for $Ca_{0.8}La_{0.2}Fe_2As_2$ at 300 K, given $\rho_{2D} = 330$ $\mu\Omega$ cm obtained from the dc limit of $\sigma_1$, $c_0 \sim 5.8$ Å is the separation of Fe-As layers, and $M$ is the number of Fermi surface sheets (which is 4 here). These materials can be considered quasi-two-dimensional systems with nearly cylindrical Fermi surfaces based on the photoemission data of Ref. [27] and Ref. [40]. Hence eq. (4) can be used to analyze charge transport in these materials. Yet another criterion for the Mott-Ioffe-Regel limit is that the mean-free path becomes comparable to the lattice constant. One can estimate the mean-free path ($l$) of the charge carriers from $l = v_F \tau$. The average Fermi velocity estimated from ARPES in La-doped $CaFe_2As_2$ is $\sim 2\times 10^6$ cm/s, which translates to a mean-free path of 2.7 Å. This mean-free path is smaller than the $a$-axis lattice constant of 3.92 Å. For the Pr-doped sample, similar calculations to those given above yield $k_F l \sim 2$ and a mean-free path of 3.5 Å, which is comparable to the lattice constant of 3.91 Å [14]. From ARPES results [27,40] the Fermi energy of the mobile carriers, i.e., occupied (unoccupied) bandwidths for electrons (holes) are between 0.02 eV and 0.07 eV in UT La- and Pr- doped $CaFe_2As_2$ which are comparable to the saturated scattering rate $\hbar/\tau$ of 0.05 eV for the former and 0.035 eV for the latter material. It is generally understood that for the quasiparticle picture in Fermi liquid theory to be applicable, $\hbar/\tau$ should be much smaller than the Fermi energy. Since $\hbar/\tau$ is similar to the Fermi energy of the carriers in the various bands, the quasiparticle picture is hardly valid for transport above 200 K.

Our observations of scattering rate saturation near the Mott-Ioffe-Regel limit that is not directly apparent in the dc resistivity in rare-earth doped $CaFe_2As_2$ are reminiscent of the findings of Hussey *et al.* in the cuprate $La_{2-x}Sr_xCuO_4$ [41]. These authors suggest that resistivity continues to increase with increasing temperature beyond the Mott-Ioffe-Regel limit because of the loss of Drude spectral weight due to dominance of electronic correlations in charge transport. It therefore follows that the iron arsenides may be considered as "bad metals". This does not contradict the observation of resistivity saturation about 600 K in the SrFe$_2$As$_2$ system because this phenomenon occurs at resistivities that are beyond the Mott-Ioffe-Regel limit of metallic transport [42].

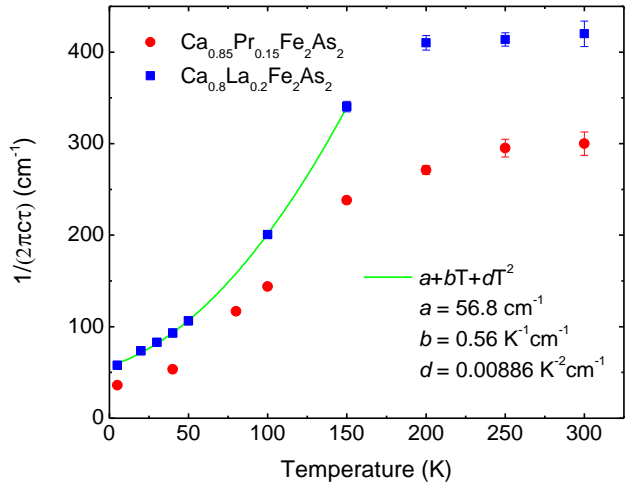

FIG. 4. Temperature dependence of the scattering rate $1/(2\pi c\tau)$ of the Drude term of Ca$_{0.8}$La$_{0.2}$Fe$_2$As$_2$ (blue squares) and Ca$_{0.85}$Pr$_{0.15}$Fe$_2$As$_2$ (red circles), and a fit for Ca$_{0.8}$La$_{0.2}$Fe$_2$As$_2$ scattering rate to the form $a+bT+dT^2$ (green line).

The temperature dependence of the scattering rate of Ca$_{0.8}$La$_{0.2}$Fe$_2$As$_2$ and Ca$_{0.85}$Pr$_{0.15}$Fe$_2$As$_2$ is shown in Fig. 4. We fit the UT Ca$_{0.8}$La$_{0.2}$Fe$_2$As$_2$ scattering rate to the form $a+bT+dT^2$. Even though the quadratic term dominates, the fit can be improved with the addition of a linear temperature-dependent term. The coefficient of the linear term "$b$" is 0.56 K$^{-1}$cm$^{-1}$. If we assume that the linear term arises from electron-phonon scattering, then the dimensionless electron-phonon coupling constant $\lambda$ can be calculated from the equation [43]:

$$\frac{\hbar}{\tau} = 2\pi\lambda k_B T. \quad (5)$$

This gives $\lambda = 0.13$, remarkably consistent with previous results that show weak electron-phonon coupling for $ab$-plane transport in the 122-iron arsenides [42].

The coefficient of the quadratic term $d = 0.008\ 86$ K$^{-2}$cm$^{-1}$ = $1.67\times10^9$ K$^{-2}$s$^{-1}$ is much larger than that of a good metal like gold (~$10^7$ K$^{-2}$s$^{-1}$) [44,45]. The temperature dependent quadratic term is likely electronic in origin and is similar to that seen in other Mott-correlated and Hund-correlated systems like V$_2$O$_3$, NdNiO$_3$, La$_{0.67}$Sr$_{0.33}$MnO$_3$, and CaRuO$_3$ [45,46]. The CT phase transition in Pr-doped CaFe$_2$As$_2$ at ~ 70 K with a hysteresis of ~ 30 K [14] precludes the preceding quantitative analysis, but we note that the temperature dependence of the scattering rate above 70 K closely resembles the data for La-doped CaFe$_2$As$_2$. However, below the CT phase transition, the normalized scattering rate of Pr-doped CaFe$_2$As$_2$ is relatively lower compared to that of the UT La-doped CaFe$_2$As$_2$. We attribute this to decreased electronic scattering upon reduction of the Fe magnetic moment in the CT phase [18].

We analyze the quadratic temperature dependence of the scattering rate with the Umklapp electron-electron scattering model of Fermi liquid theory [47]:

$$\frac{\hbar}{\tau} = A\frac{(k_B T)^2}{E_F}. \quad (6)$$

We estimate the dimensionless constant $A \approx 4$ assuming an average Fermi energy $E_F \approx 30$ meV in La-doped CaFe$_2$As$_2$. This value of $A$ is somewhat larger than that obtained for Co-doped BaFe$_2$As$_2$ in Ref. [47] indicating comparatively enhanced effective Umklapp scattering in rare-earth-doped CaFe$_2$As$_2$. A quadratic temperature dependence of the scattering rate has been seen before in Co-doped BaFe$_2$As$_2$ up to room temperature without saturation [47,48], and this is likely due to its larger Fermi energy. We expect the scattering rate to saturate in Co-doped BaFe$_2$As$_2$ if heated above room temperature. Clearly, even higher temperatures are required for attaining the Mott-Ioffe-Regel limit in conventional metals that possess larger Fermi energy [39]. Saturation of scattering rate has been observed by infrared spectroscopy in the iron chalcogenide FeTe$_{0.55}$Se$_{0.45}$, a system with a low Fermi energy and strong electronic correlations [49]. In the La-doped CaFe$_2$As$_2$ we see a crossover from a predominantly quadratic temperature-dependent scattering rate below 150 K indicating the presence of coherent, mobile charges to saturation of the scattering rate above 200 K associated with incoherent transport. It appears that the main reason for the saturation of the scattering rate in the rare-earth-doped CaFe$_2$As$_2$ systems is enhanced electron-electron scattering that increases with temperature leading to a breakdown of the quasiparticle picture. The large scattering rate is due to a combination of reasons: low Fermi energy of charge carriers; both normal and Umklapp scattering events between electrons and holes contributing to enhanced dissipation; and coherent carriers scattering off incoherent charges. At low temperatures, where the quasiparticle concept may be valid as exemplified by eq. (6), there is significant spectral weight in the overdamped Lorentz oscillator (see the oscillators labeled Lorentzian 1 in Appendix C). Some of this spectral weight is due to incoherent and localized charges that coexist with mobile charges. Moreover, an increasing number of mobile charges become incoherent with increasing temperature, as seen by the decrease of Drude spectral weight with increasing temperature, and that this spectral weight is recovered at an energy scale of ~ 0.5 eV which is much larger than the Fermi energies of the electrons and holes. Taken together, the observations in our work make it difficult to classify the rare-earth CaFe$_2$As$_2$ system as a conventional Fermi liquid.

In order to confirm the results of the preceding analysis based on fits to the Drude-Lorentz model, we perform the extended Drude model analysis to examine the frequency dependence of scattering rate. Here we use the form [50]:

$$\frac{1}{\tau(\omega)} = -\frac{\omega_p^2}{\omega}\text{Im}\left(\frac{1}{\tilde{\varepsilon}(\omega)-\varepsilon_H}\right), \quad (7)$$

where $\omega_p^2$ is calculated from the integral of $\sigma_1$ up to 500 cm$^{-1}$, $\tilde{\varepsilon}(\omega)$ is the complex dielectric function and $\varepsilon_H$ represents the contribution of higher-energy interband transitions. Note that the choice of upper frequency cutoff in the integral used for calculating $\omega_p^2$ does not affect the frequency dependence of the scattering rate. Figures 5(a) and 5(b) show frequency-dependent scattering rate of La- and Pr-doped CaFe$_2$As$_2$, respectively, for representative temperatures. At high temperatures [like 200 K for Ca$_{0.8}$La$_{0.2}$Fe$_2$As$_2$ shown in Fig. 5(a)], the scattering rate hardly shows frequency dependence, which is consistent with saturation of scattering rate as a function of temperature that is extracted from fits of the conductivity to a one Drude-one Lorentz model. Low-temperature scattering rate follows a quadratic form $C+B\omega^2$ [51], which gives similar coefficient B for both samples. A linear frequency-dependent term is not included because it does not improve the fits. Such a term may be relevant at frequencies below the 60-cm$^{-1}$ lower cutoff of our data. According to Ref. [51], the upper cutoff frequency for the quadratic fit at each temperature is determined by noting that $\hbar\omega$ should be smaller or comparable to $2\pi k_B T$. We also note that the temperature dependence of the low frequency limit for $1/\tau(\omega,T)$ based on the extended Drude model is essentially the same as the temperature dependence of the scattering rate obtained from the Drude-Lorentz model and plotted in Fig. 4. If we compare the coefficients of the temperature-dependent quadratic term (from Drude-Lorentz analysis and extended Drude analysis) and the frequency-dependent quadratic term (from extended Drude analysis) of the scattering rate in La-doped CaFe$_2$As$_2$, and use the scattering rate form [51,52]:

$$\frac{1}{\tau}(\omega,T) \propto A_0[(\hbar\omega)^2 + (p\pi k_B T)^2], \quad (8)$$

we get $p = 1.53$. This value of $p$ is very close to the value obtained in BaFe$_{1.8}$Co$_{0.2}$As$_2$ and underdoped cuprates [52,53]. The value of $p$ should be 2 for a conventional Fermi liquid.

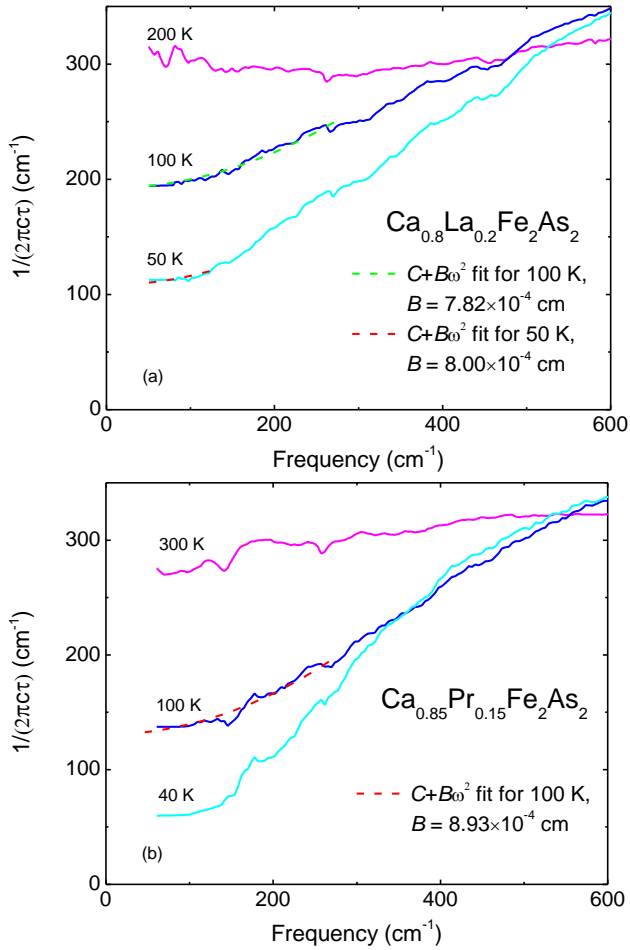

FIG. 5. Frequency dependent scattering rate of (a) La- and (b) Pr-doped CaFe$_2$As$_2$. Scattering rate of both samples shows saturation at high temperatures (scattering rate is flat and frequency independent). For temperatures $\leq$ 100 K, the quadratic term coefficient $B$ is temperature independent in La-doped CaFe$_2$As$_2$ and is similar in magnitude to that in the Pr-doped sample in the UT phase at 100 K. However, in the Pr-doped sample in the CT phase (40 K), the scattering rate curve is clearly different from that in the UT phase (100 K), which indicates reconstruction of the Fermi surface.

## C. Interband transitions

Next we discuss the physical interpretation of the Lorentz oscillators that represent interband transitions. Unlike the UT phase, a Lorentz oscillator is required to fit the hump in $\sigma_1$ about 400 cm$^{-1}$ in the CT phase in Ca$_{0.85}$Pr$_{0.15}$Fe$_2$As$_2$, as shown in Fig. 10 in Appendix C. According to Ref. [27], at zone center, the $\beta$ band shifts down below Fermi energy across CT phase transition, leaving the $\alpha$ band still above the Fermi energy. The gap between the top of the two bands is about 30 meV (240 cm$^{-1}$) at the $\Gamma$ point and the gap increases at larger wavevectors. So we may conclude that the hump in conductivity 400 cm$^{-1}$ is from the interband transition between $\alpha$ and $\beta$ band in the CT phase. The optical transition between the weakly hybridized Fe-$d$ and As-$p$ band to an unoccupied Fe-$d$ band [54] is centered about 7000 cm$^{-1}$ for Pr-doped CaFe$_2$As$_2$ [see Fig. 1(b)]. The center frequency of this interband transition after structure collapse increases by about 500 cm$^{-1}$ which we also attribute primarily to the downward shift of the $\beta$ band.

## IV. SUMMARY

In summary, we have obtained the frequency- and temperature-dependent *ab*-plane optical conductivity of crystals of rare-earth-doped $CaFe_2As_2$. For UT La-doped and Pr-doped $CaFe_2As_2$, the scattering rate reveals a dominant scattering channel quadratic in temperature and frequency. We also find saturation of the scattering rate above 200 K near the Mott-Ioffe-Regel limit in UT La-doped and Pr-doped $CaFe_2As_2$. The spectral weight of free charge carriers in the UT phase decreases with increasing temperature in a broad temperature range and is recovered at an energy scale of ~ 0.5 eV which is much larger than the Fermi energy scale. Given that the phenomena we observe in rare-earth-doped $CaFe_2As_2$ are similar to that seen in other correlated metals, we are forced to conclude that the dominant scattering mechanism is of electronic origin, and these materials are not canonical Fermi liquids. Below the CT phase transition in Pr-doped $CaFe_2As_2$, we observe a decrease of the scattering rate due to weakening of electronic correlations, and a decrease in mobile carrier density which is consistent with the partial loss of the hole Fermi surfaces.

## ACKNOWLEDGMENTS

M.M.Q acknowledges financial support from the Virginia Space Grant Consortium for this project. Work at the University of Maryland was supported by AFOSR through Grant No. FA9550-14-1-0332. Z.X. acknowledges the use of REFFIT software (http://optics.unige.ch/alexey/reffit.html) for Drude-Lorentz fits to the optical conductivity data.

## APPENDIX A: REFLECTANCE, ELLIPSOMETRY, AND DATA ANALYSIS TO OBTAIN *ab*-PLANE OPTICAL CONSTANTS

Fig. 6(a) and 6(b) show *ab*-plane reflectance spectra of $Ca_{0.8}La_{0.2}Fe_2As_2$ and $Ca_{0.85}Pr_{0.15}Fe_2As_2$ respectively. The rather high reflectance at low frequencies is clearly indicative of metallicity. There is no evidence of bulk superconductivity in the infrared reflectance. This is consistent with the report of very low volume fraction superconductivity in these materials [14]. The reflectance of the $Ca_{0.85}Pr_{0.15}Fe_2As_2$ crystal in the far- and midinfrared region shows subtle changes across the structure collapse transition which are more obvious in the optical conductivity, as discussed in the main text. The reflectance spectra were obtained in the Bruker Vertex 80v FTIR spectrometer that is fitted with an ultrahigh-vacuum chamber designed in-house for use with a continuous-flow liquid helium cryostat.

In single crystal samples, the absolute value of the dc resistivity occasionally has a systematic error due to the difficulty in precise measurements of the geometry of the crystals. We use Hagen-Rubens extrapolation of room temperature infrared reflectance to determine the absolute value of the room temperature dc conductivity (in Hagen-Rubens extrapolation, dc conductivity is the only fit parameter). Then relative dc resistivity data measured at lower temperatures (in Ref. [14]) are used to find absolute temperature-dependent dc conductivities which are employed in Hagen-Rubens extrapolations of temperature-dependent infrared reflectance for Kramers-Kronig analysis. Hence, the dc extension of the optical conductivity agrees well with measured dc conductivity.

Frequency and temperature dependence of the ellipsometric coefficients $\Psi$ and $\Delta$ for $Ca_{0.8}La_{0.2}Fe_2As_2$ and $Ca_{0.85}Pr_{0.15}Fe_2As_2$ are shown in Fig. 7. The ellipsometry data shown in Fig. 7 was obtained in an ultrahigh-vacuum chamber designed in-house for use with a continuous-flow liquid helium cryostat and the Woollam VASE instrument. The pseudodielectric function calculated from the ellipsometric coefficients of the rare-earth-doped $CaFe_2As_2$ crystals with the $c$ axis normal to the crystal surface is nearly the same as the $ab$-plane optical constants. This is unlike the superconducting cuprates in which $c$-axis optical constants are quite different from $ab$-plane ones [55,56], and the pseudodielectric function for crystals with $c$ axis normal to the sample surface has to be corrected to obtain the $ab$-plane optical constants. In the 122 iron arsenides the $ab$-plane and $c$-axis optical conductivities differ by 20% − 30% [57,58]. In the absence of $c$-axis optical spectroscopy data, it is reasonable for us to assume a similar level of anisotropy in the rare-earth-doped $CaFe_2As_2$. According to Jellison and Baba [59], for the special case like the measurements in principle symmetry directions (optical axis, i.e., $c$ axis is perpendicular to the sample surface), the complex pseudodielectric function $\langle\varepsilon\rangle = \langle\varepsilon_1\rangle - i\langle\varepsilon_2\rangle$ measured directly from ellipsometry data can be expressed in terms of $\varepsilon_{ab}$ and $\varepsilon_c$:

$$\langle\varepsilon\rangle = sin^2\varphi\left[1 + sin^2\varphi\left(\frac{\varepsilon_{ab}(\varepsilon_{ab}-sin^2\varphi)^{\frac{1}{2}} - \left[\frac{\varepsilon_{ab}(\varepsilon_c-sin^2\varphi)}{\varepsilon_c}\right]^{\frac{1}{2}}}{\varepsilon_{ab}(1-sin^2\varphi) - (\varepsilon_{ab}-sin^2\varphi)^{\frac{1}{2}}\left[\frac{\varepsilon_{ab}(\varepsilon_c-sin^2\varphi)}{\varepsilon_c}\right]^{\frac{1}{2}}}\right)^2\right], \quad (A1)$$

where $\varphi$ is the angle between the beam and surface normal, $\varepsilon_{ab}$ and $\varepsilon_c$ are the $ab$-plane and $c$-axis complex dielectric functions respectively. The pseudodielectric function can be expressed by Taylor expansion in powers of $\Delta\varepsilon = \varepsilon_c - \varepsilon_{ab}$ (we keep three terms here):

$$\langle\varepsilon\rangle \approx \varepsilon_{ab} - \frac{\Delta\varepsilon}{\varepsilon_{ab}-1}$$
$$+ \frac{\Delta\varepsilon^2}{4\varepsilon_{ab}(\varepsilon_{ab}-1)^2}\frac{4\varepsilon_{ab}^2 - \varepsilon_{ab} - 3\varepsilon_{ab}sin^2\varphi + sin^2\varphi}{\varepsilon_{ab} - sin^2\varphi}. \quad (A2)$$

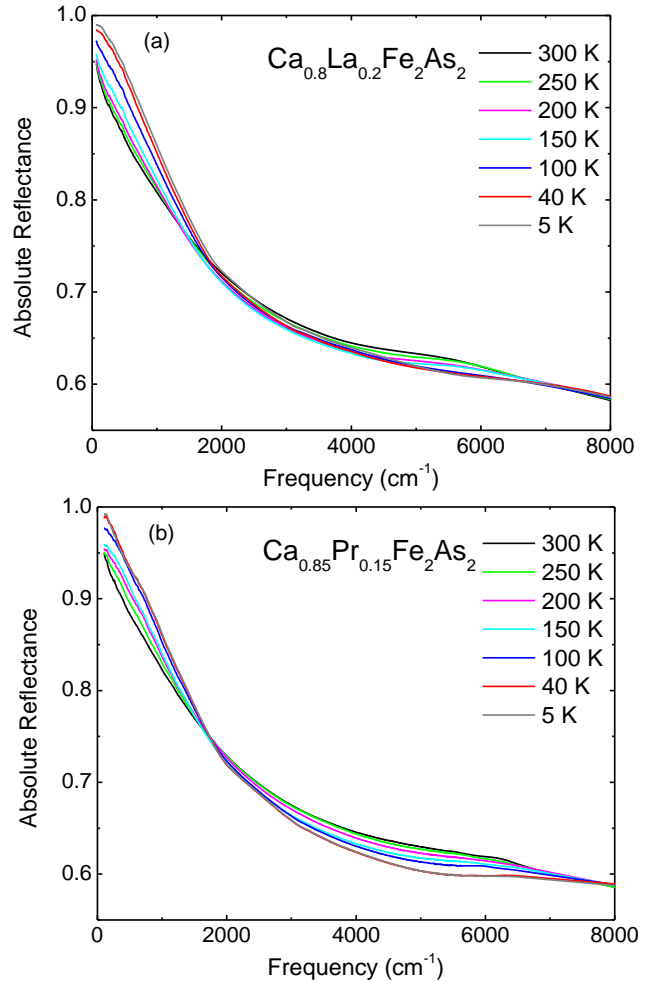

FIG. 6. Frequency dependence of absolute reflectance at representative temperatures for (a) $Ca_{0.8}La_{0.2}Fe_2As_2$ and (b) $Ca_{0.85}Pr_{0.15}Fe_2As_2$.

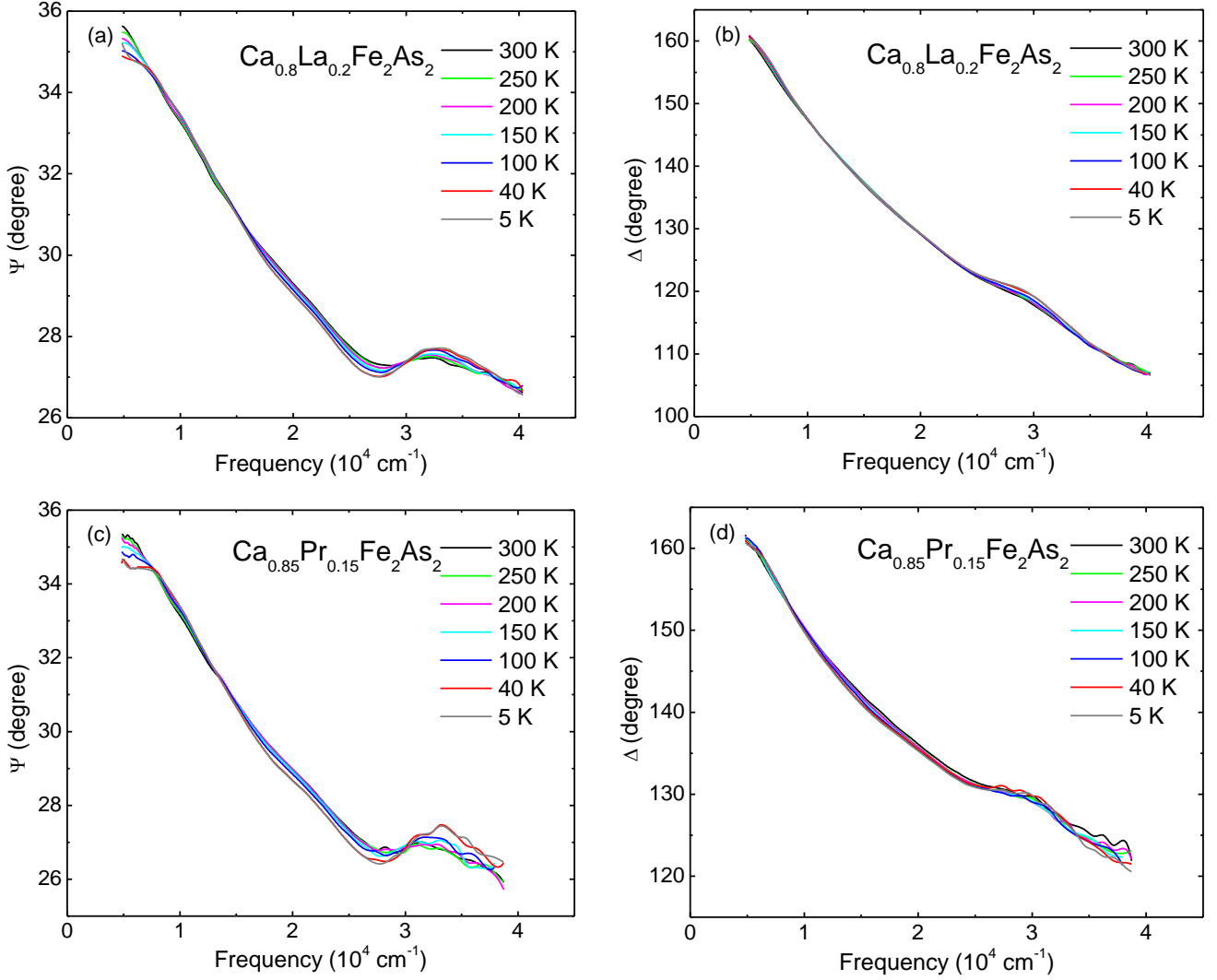

FIG. 7. (a, b) Frequency- and temperature-dependent ellipsometric coefficients $\Psi$ and $\Delta$ for $Ca_{0.8}La_{0.2}Fe_2As_2$; (c, d) frequency- and temperature-dependent ellipsometric coefficients $\Psi$ and $\Delta$ for $Ca_{0.85}Pr_{0.15}Fe_2As_2$.

Our assumption is that, $\Delta\varepsilon / (\varepsilon_{ab} -1) \sim 20\% - 30\%$, i.e., $0.2 - 0.3$, so the third term of eq. (A2) which depends on the angle of incidence should be quite small (less than 1% for $\langle\varepsilon_2\rangle$), and this is confirmed from our multiple angle of incidence ellipsometry measurements (as shown in Fig. 8). The pseudodielectric function we measured hardly shows any angle of incidence dependence. At 15 000 cm$^{-1}$, $\langle\varepsilon_2\rangle$ is about 10, which makes the contribution to $|\langle\varepsilon\rangle|$ of the term $\Delta\varepsilon / (\varepsilon_{ab} -1)$ about $2 - 3\%$ at most. Also when $\langle\varepsilon_1\rangle$ is small, both reflectance and phase used in Kramers-Kronig analysis based on Ref. [30] are mainly determined by $\langle\varepsilon_2\rangle$. Above 15 000 cm$^{-1}$ to highest measured frequencies, the uncertainty in $ab$-plane $\varepsilon_2$ may be between 2% and 10% due to possible contribution to $\langle\varepsilon_2\rangle$ from $c$-axis optical properties. However, this has negligibly small effect on calculations of $ab$ plane optical constants below 6000 cm$^{-1}$. Thus we can say $\langle\varepsilon\rangle \approx \varepsilon_{ab}$, i.e., the pseudodielectric function is the *ab*-plane dielectric function within the uncertainties stated above. In fact, the larger $\varepsilon_{ab}$ the smaller the influence of the *c*-axis optical constants on the pseudodielectric function. So below 20 000 cm$^{-1}$ (where $\langle \varepsilon_2 \rangle$ is quite large), the pseudodielectric function we get directly from ellipsometry data is an accurate representation of the *ab*-plane dielectric function (within 3% uncertainty for $\varepsilon_2$), and correction due to *c*-axis optical properties is not necessary. Another piece of supporting evidence is that between 4800 cm$^{-1}$ and 6400 cm$^{-1}$ (0.6 – 0.8 eV) the *ab*-plane absolute reflectance we measured is remarkably consistent to within 0.5% of the reflectance generated from pseudodielectric function. For the purpose of performing Kramers-Kronig analysis on the infrared reflectance constrained by ellipsometry data, we assume the reflectance generated from the ellipsometry data is more reliable (random uncertainty in reflectance generated from ellipsometric coefficients is about 0.2%.) Next we adjust the *ab*-plane infrared reflectance in the range 4800 – 6000 cm$^{-1}$ to match the reflectance generated from ellipsometric coefficients. The reflectance uncertainty in the range 4800 – 6000 cm$^{-1}$ is around 0.5%, which leads to 1.5% uncertainty in conductivity in the same frequency range and even lower uncertainty of about 1% in the far infrared region. To summarize, the *ab*-plane optical conductivity below 6000 cm$^{-1}$ we obtain from this method has a few percent systematic error at most, and the relative uncertainty for different temperatures is much smaller.

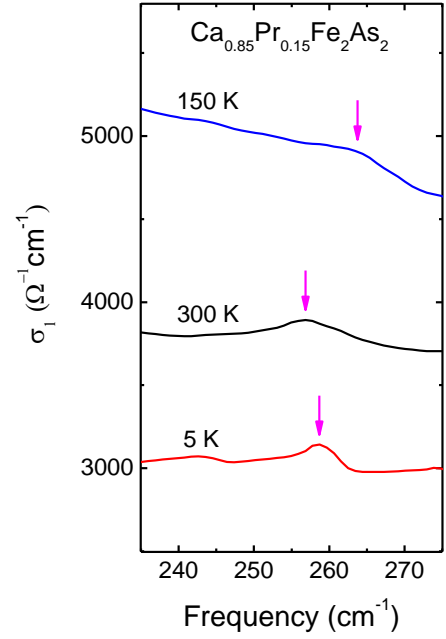

FIG. 9. Temperature dependence of the Fe-As phonon feature in the optical conductivity of Ca$_{0.85}$Pr$_{0.15}$Fe$_2$As$_2$. Arrows indicate the center frequencies of the phonon.

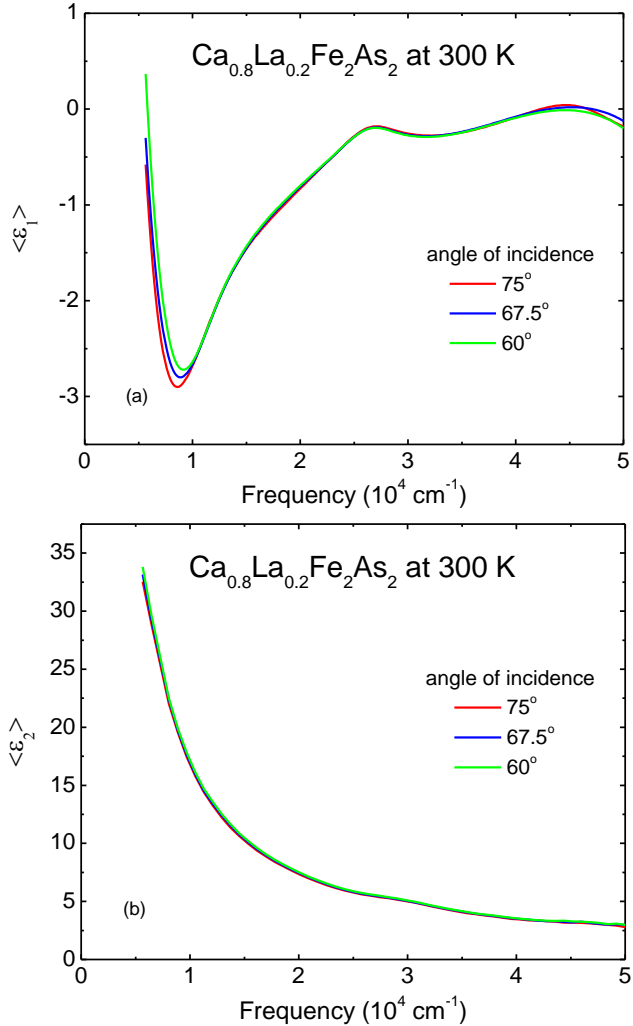

FIG. 8. (a, b) Real and imaginary parts ($\langle\varepsilon_1\rangle$ and $\langle\varepsilon_2\rangle$) of the pseudodielectric function of Ca$_{0.8}$La$_{0.2}$Fe$_2$As$_2$ at room temperature for different angles of incidence.

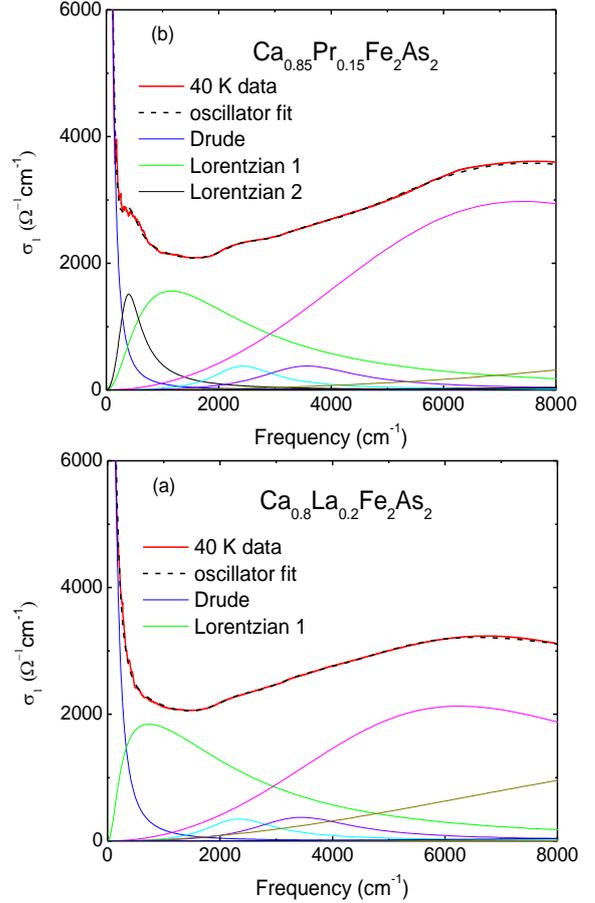

FIG. 10. Oscillator fits to the measured $\sigma_1$ at 40 K for (a) Ca$_{0.8}$La$_{0.2}$Fe$_2$As$_2$ and (b) Ca$_{0.85}$Pr$_{0.15}$Fe$_2$As$_2$. The thick solid line (red) is the data and the black dashed line is the sum of the Drude-Lorentz oscillators. The Drude and Lorentz oscillators used in the fits are shown as thin solid lines.

## APPENDIX B: PHONON SHIFT ACROSS CT PHASE TRANSITION

Here we discuss the effect of the CT phase transition in $Ca_{0.85}Pr_{0.15}Fe_2As_2$ on the optical phonons. For the parent compound (space-group I4/mmm) $CaFe_2As_2$, there are two *ab*-plane infrared-active $E_u$ modes [60,61]. Both phonons have been observed in Pr-doped $CaFe_2As_2$, although the impact of the structural transition is more clearly evident in the behavior of the higher-frequency Fe-As vibration. Figure 9 shows the impact of structural collapse on the center frequency of the Fe-As phonon in $Ca_{0.85}Pr_{0.15}Fe_2As_2$. The phonon center frequency in the UT phase increases as temperature decreases, as in the 300-K and 150-K data. However, the center frequency of Fe-As phonon decreases below CT phase transition as seen in the 5-K data. This is direct evidence of the CT phase transition from infrared spectroscopy. We note that the feature at ~ 140 cm$^{-1}$ in the conductivity of Ca$_{0.85}$Pr$_{0.15}$Fe$_2$As$_2$ [Fig. 1(b)] is attributed to the phonon mode associated primarily with vibrations of the Ca ion [61]. We do not expect Pr doping to significantly affect this phonon feature because the ionic radius of the Pr ion is nearly the same as that of the Ca ion. Also, this phonon feature becomes weaker and possibly moves to ~ 175 cm$^{-1}$ across the structural collapse into the CT phase [Fig. 1(b)]. This phonon feature is much weaker in the conductivity data on Ca$_{0.8}$La$_{0.2}$Fe$_2$As$_2$. It is likely broadened out due to the lower concentration of Ca and the significant difference in the ionic radii of the Ca and La ions [14].

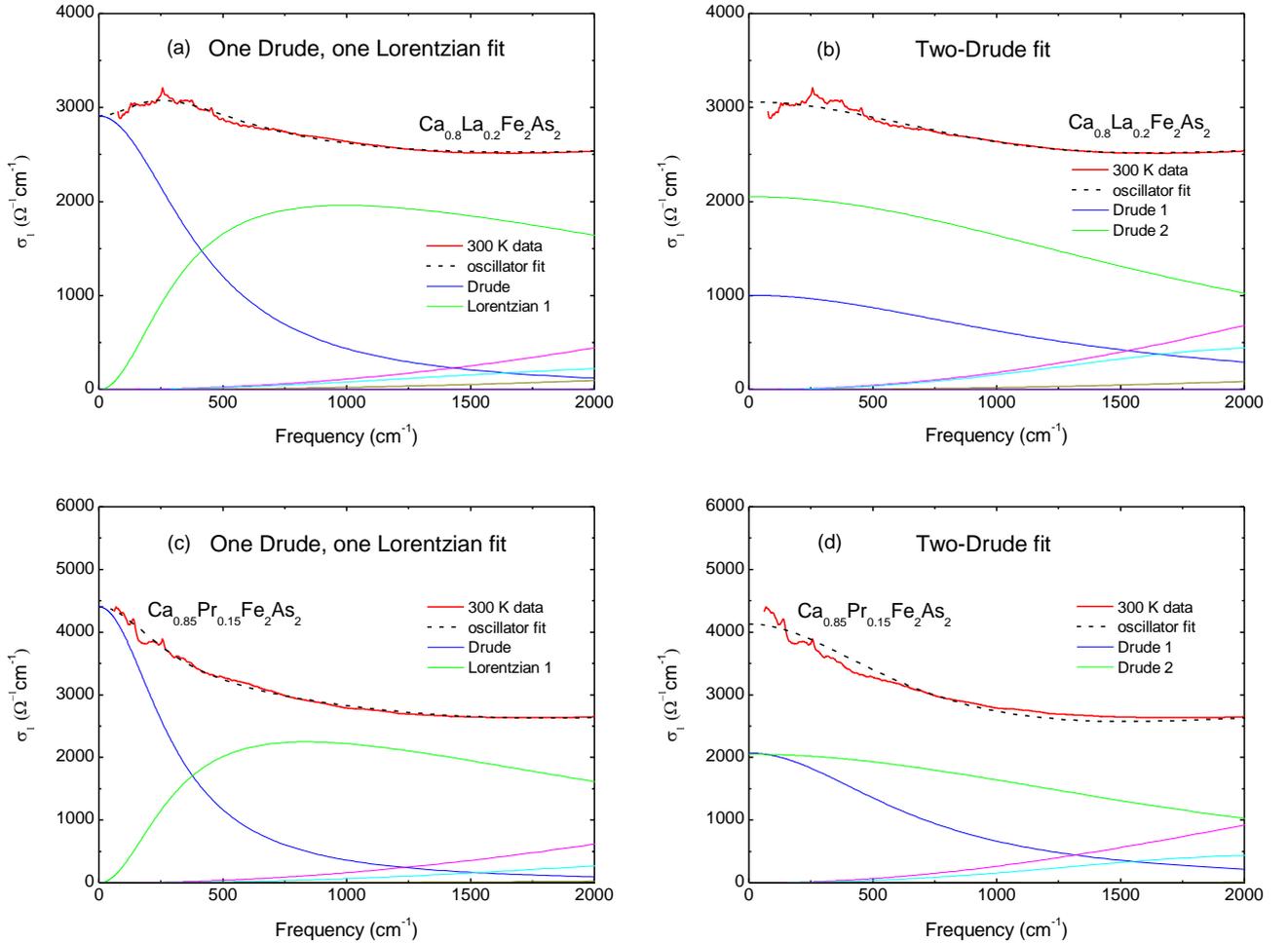

FIG. 11. (a) One Drude-one Lorentzian fit and (b) two-Drude fit to room temperature infrared conductivity of $Ca_{0.8}La_{0.2}Fe_2As_2$; (c) One Drude-one Lorentzian fit and (d) two-Drude fit to room temperature infrared conductivity of $Ca_{0.85}Pr_{0.15}Fe_2As_2$. The low frequency tails of higher-lying Lorentz oscillators can also be seen as thin solid lines in all panels.

## APPENDIX C: REPRESENTATIVE FITS

Both the real and imaginary parts of the conductivity are fit well to the Drude-Lorentz model. Here we show and discuss the fits to the real part of the conductivity ($\sigma_1$). Figure 10 shows a comparison of the fits to $\sigma_1$ at 40 K for Ca$_{0.8}$La$_{0.2}$Fe$_2$As$_2$ (UT phase) and Ca$_{0.85}$Pr$_{0.15}$Fe$_2$As$_2$ (CT phase). In the UT phase, one Drude mode and one Lorentz oscillator is sufficient for a good fit to the low-frequency optical conductivity. Unlike the UT phase, another Lorentz oscillator (Lorentzian 2) is required to fit the hump in $\sigma_1$ around 400 cm$^{-1}$ in the CT phase. The error bars of Drude parameters plotted in Fig. 3 and 4 in the main text are calculated as follows. We manually vary each Drude parameter, while fitting the other parameters of the Drude and Lorentz modes, until the sum of the squared error between data and model increases by 10% of the best fit value.

The one Drude-one Lorentzian fit and two-Drude fit (one broad and one narrow) to the low-frequency, room temperature conductivity of La-doped and Pr-doped CaFe$_2$As$_2$ are shown in Fig. 11. In both materials, there is clear discrepancy between the data and the two-Drude fits for frequencies below 700 cm$^{-1}$. Especially for Ca$_{0.8}$La$_{0.2}$Fe$_2$As$_2$, due to the decreasing conductivity at very low frequencies, the two-Drude fits are emphatically ruled out. On the other hand, the one Drude-one Lorentzian fits work well for the infrared data in the UT phase at room temperature and at all measured cryogenic temperatures.

[1] G. R. Stewart, Rev. Mod. Phys. **83**, 1589 (2011).

[2] J. Paglione and R. L. Greene, Nat. Phys. **6**, 645 (2010).

[3] D. C. Johnston, Adv. Phys. **59**, 803 (2010).

[4] T. Park, E. Park, H. Lee, T. Klimczuk, E. D. Bauer, F. Ronning, and J. D. Thompson, J. Phys. Condens. Matter **20**, 322204 (2008).

[5] M. S. Torikachvili, S. L. Bud'ko, N. Ni, and P. C. Canfield, Phys. Rev. Lett. **101**, 057006 (2008).

[6] W. Yu, A. A. Aczel, T. J. Williams, S. L. Bud'ko, N. Ni, P. C. Canfield, and G. M. Luke, Phys. Rev. B **79**, 020511 (2009).

[7] A. I. Goldman, A. Kreyssig, K. Prokeš, D. K. Pratt, D. N. Argyriou, J. W. Lynn, S. Nandi, S. A. J. Kimber, Y. Chen, Y. B. Lee, G. Samolyuk, J. B. Leão, S. J. Poulton, S. L. Bud'ko, N. Ni, P. C. Canfield, B. N. Harmon, and R. J. McQueeney, Phys. Rev. B **79**, 024513 (2009).

[8] A. Kreyssig, M. A. Green, Y. Lee, G. D. Samolyuk, P. Zajdel, J. W. Lynn, S. L. Bud'ko, M. S. Torikachvili, N. Ni, S. Nandi, J. B. Leão, S. J. Poulton, D. N. Argyriou, B. N. Harmon, R. J. McQueeney, P. C. Canfield, and A. I. Goldman, Phys. Rev. B **78**, 184517 (2008).

[9] P. L. Alireza, Y. T. C. Ko, J. Gillett, C. M. Petrone, J. M. Cole, G. G. Lonzarich, and S. E. Sebastian, J. Phys. Condens. Matter **21**, 012208 (2009).

[10] K. Matsubayashi, N. Katayama, K. Ohgushi, A. Yamada, K. Munakata, T. Matsumoto, and Y. Uwatoko, J. Phys. Soc. Japan **78**, 073706 (2009).

[11] R. Mittal, S. K. Mishra, S. L. Chaplot, S. V. Ovsyannikov, E. Greenberg, D. M. Trots, L. Dubrovinsky, Y. Su, T. Brueckel, S. Matsuishi, H. Hosono, and G. Garbarino, Phys. Rev. B **83**, 054503 (2011).

[12] W. Uhoya, A. Stemshorn, G. Tsoi, Y. K. Vohra, A. S. Sefat, B. C. Sales, K. M. Hope, and S. T. Weir, Phys. Rev. B **82**, 144118 (2010).

[13] T. Yildirim, Phys. Rev. Lett. **102**, 037003 (2009).

[14] S. R. Saha, N. P. Butch, T. Drye, J. Magill, S. Ziemak, K. Kirshenbaum, P. Y. Zavalij, J. W. Lynn, and J. Paglione, Phys. Rev. B **85**, 024525 (2012).

[15] M. Danura, K. Kudo, Y. Oshiro, S. Araki, T. C. Kobayashi, and M. Nohara, J. Phys. Soc. Jpn **80**, 10371 (2011).

[16] S. Kasahara, T. Shibauchi, K. Hashimoto, Y. Nakai, H. Ikeda, T. Terashima, and Y. Matsuda, Phys. Rev. B **83**, 060505 (2011).

[17] K. Gofryk, M. Pan, C. Cantoni, B. Saparov, J. E. Mitchell, and A. S. Sefat, Phys. Rev. Lett. **112**, 047005 (2014).

[18] H. Gretarsson, S. R. Saha, T. Drye, J. Paglione, J. Kim, D. Casa, T. Gog, W. Wu, S. R. Julian, and Y.-J. Kim, Phys. Rev. Lett. **110**, 047003 (2013).

[19] L. Ma, G.-F. Ji, J. Dai, S. R. Saha, T. Drye, J. Paglione, and W.-Q. Yu, Chinese Phys. B **22**, 057401 (2013).

[20] A. Sanna, G. Profeta, S. Massidda, and E. K. U. Gross, Phys. Rev. B **86**, 014507 (2012).

[21] B. Lv, L. Deng, M. Gooch, F. Wei, Y. Sun, J. K. Meen, Y.-Y. Xue, B. Lorenz, and C.-W. Chu, Proc. Natl. Acad. Sci. U. S. A. **108**, 15705 (2011).

[22] D. K. Pratt, Y. Zhao, S. A. J. Kimber, A. Hiess, D. N. Argyriou, C. Broholm, A. Kreyssig, S. Nandi, S. L. Bud'ko, N. Ni, P. C. Canfield, R. J. McQueeney, and A. I. Goldman, Phys. Rev. B **79**, 060510 (2009).

[23] S. Mandal, R. E. Cohen, and K. Haule, Phys. Rev. B **90**, 060501 (2014).

[24] K. Gofryk, B. Saparov, T. Durakiewicz, A. Chikina, S. Danzenbächer, D. V. Vyalikh, M. J. Graf, and A. S. Sefat, Phys. Rev. Lett. **112**, 186401 (2014).

[25] R. S. Dhaka, R. Jiang, S. Ran, S. L. Bud'ko, P. C. Canfield, B. N. Harmon, A. Kaminski, M. Tomić, R. Valentí, and Y. Lee, Phys. Rev. B **89**, 020511 (2014).

[26] Y.-Z. Zhang, H. C. Kandpal, I. Opahle, H. O. Jeschke, and R. Valentí, Phys. Rev. B **80**, 094530 (2009).

[27] D. F. Xu, D. W. Shen, J. Jiang, Z. R. Ye, X. Liu, X. H. Niu, H. C. Xu, Y. J. Yan, T. Zhang, B. P. Xie, and D. L. Feng, Phys. Rev. B **90**, 214519 (2014).

[28] R. Yang, C. Le, L. Zhang, B. Xu, W. Zhang, K. Nadeem, H. Xiao, J. Hu, and X. Qiu, Phys. Rev. B **91**, 224507 (2015).

[29] C. C. Homes, M. Reedyk, D. A. Cradles, and T. Timusk, Appl. Opt. **32**, 2976 (1993).

[30] I. Bozovic, Phys. Rev. B **42**, 1969 (1990).

[31] M. Dressel and G. Grüner, *Electrodynamics of Solids* (Cambridge University Press, Cambridge, 2002).

[32] K. Takenaka, J. Nohara, R. Shiozaki, and S. Sugai, Phys. Rev. B **68**, 134501 (2003).

[33] W. Z. Hu, J. Dong, G. Li, Z. Li, P. Zheng, G. F. Chen, J. L. Luo, and N. L. Wang, Phys. Rev. Lett. **101**, 257005 (2008).

[34] J. J. Tu, J. Li, W. Liu, A. Punnoose, Y. Gong, Y. H. Ren, L. J. Li, G. H. Cao, Z. A. Xu, and C. C. Homes, Phys. Rev. B **82**, 174509 (2010).

[35] D. Wu, N. Barišić, P. Kallina, A. Faridian, B. Gorshunov, N. Drichko, L. J. Li, X. Lin, G. H. Cao, Z. A. Xu, N. L. Wang, and M. Dressel, Phys. Rev. B **81**, 100512 (2010).

[36] M. Nakajima, S. Ishida, K. Kihou, Y. Tomioka, T. Ito, Y. Yoshida, C. H. Lee, H. Kito, A. Iyo, H. Eisaki, K. M. Kojima, and S. Uchida, Phys. Rev. B **81**, 104528 (2010).

[37] B. Cheng, B. F. Hu, R. Y. Chen, G. Xu, P. Zheng, J. L. Luo, and N. L. Wang, Phys. Rev. B **86**, 134503 (2012).

[38] X. B. Wang, H. P. Wang, T. Dong, R. Y. Chen, and N. L. Wang, Phys. Rev. B **90**, 144513 (2014).

[39] O. Gunnarsson, M. Calandra, and J. E. Han, Rev. Mod. Phys. **75**, 1085 (2003).

[40] Y.-B. Huang, P. Richard, J.-H. Wang, X.-P. Wang, X. Shi, N. Xu, Z. Wu, A. Li, J.-X. Yin, T. Qian, B. Lv, C.-W. Chu, S.-H. Pan, M. Shi, and H. Ding, Chinese Phys. Lett. **30**, 017402 (2013).

[41] N. E. Hussey, K. Takenaka, and H. Takagi, Philos. Mag. **84**, 2847 (2004).

[42] P. L. Bach, S. R. Saha, K. Kirshenbaum, J. Paglione, and R. L. Greene, Phys. Rev. B **83**, 212506 (2011).

[43] P. B. Allen, W. E. Pickett, and H. Krakauer, Phys. Rev. B **37**, 7482 (1988).

[44] G. R. Parkins, W. E. Lawrence, and R. W. Christy, Phys. Rev. B **23**, 6408 (1981).

[45] P. Xu, T. J. Huffman, N. C. Branagan, M. M. Qazilbash, P. Srivastava, T. Goehringer, G. Yong, V. Smolyaninova, and R. Kolagani, Philos. Mag. **95**, 2078 (2015).

[46] X. Deng, A. Sternbach, K. Haule, D. N. Basov, and G. Kotliar, Phys. Rev. Lett. **113**, 246404 (2014).

[47] N. Barišić, D. Wu, M. Dressel, L. J. Li, G. H. Cao, and Z. A. Xu, Phys. Rev. B **82**, 054518 (2010).

[48] F. Rullier-Albenque, D. Colson, A. Forget, and H. Alloul, Phys. Rev. Lett. **103**, 057001 (2009).

[49] C. C. Homes, A. Akrap, J. S. Wen, Z. J. Xu, Z. W. Lin, Q. Li, and G. D. Gu, Phys. Rev. B **81**, 180508 (2010).

[50] M. M. Qazilbash, J. J. Hamlin, R. E. Baumbach, L. Zhang, D. J. Singh, M. B. Maple, and D. N. Basov, Nat. Phys. **5**, 647 (2009).

[51] C. Berthod, J. Mravlje, X. Deng, R. Žitko, D. van der Marel, and A. Georges, Phys. Rev. B **87**, 115109 (2013).

[52] A. Tytarenko, Y. Huang, A. de Visser, S. Johnston, and E. van Heumen, Sci. Rep. **5**, 12421 (2015).

[53] S. I. Mirzaei, D. Stricker, J. N. Hancock, C. Berthod, A. Georges, E. van Heumen, M. K. Chan,

X. Zhao, Y. Li, M. Greven, N. Barišić, and D. van der Marel, Proc. Natl. Acad. Sci. U. S. A. **110**, 5774 (2013).

[54] A. Charnukha, P. Popovich, Y. Matiks, D. L. Sun, C. T. Lin, A. N. Yaresko, B. Keimer, and A. V. Boris, Nat. Commun. **2**, 219 (2011).

[55] S. L. Cooper, D. Reznik, A. Kotz, M. A. Karlow, R. Liu, M. V. Klein, W. C. Lee, J. Giapintzakis, D. M. Ginsberg, B. W. Veal, and A. P. Paulikas, Phys. Rev. B **47**, 8233 (1993).

[56] D. N. Basov, S. I. Woods, A. S. Katz, E. J. Singley, R. C. Dynes, M. Xu, D. G. Hinks, C. C. Homes, and M. Strongin, Science. **283**, 49 (1999).

[57] B. Cheng, Z. G. Chen, C. L. Zhang, R. H. Ruan, T. Dong, B. F. Hu, W. T. Guo, S. S. Miao, P. Zheng, J. L. Luo, G. Xu, P. Dai, and N. L. Wang, Phys. Rev. B **83**, 144522 (2011).

[58] Z. G. Chen, T. Dong, R. H. Ruan, B. F. Hu, B. Cheng, W. Z. Hu, P. Zheng, Z. Fang, X. Dai, and N. L. Wang, Phys. Rev. Lett. **105**, 097003 (2010).

[59] J. Jellison and J. S. Baba, J. Opt. Soc. Am. A **23**, 468 (2006).

[60] M. Rende, Y. Li, Z. Bai, L. Wang, and L. Chen, Phys. B Condens. Matter **405**, 4226 (2010).

[61] A. Charnukha, D. Pröpper, T. I. Larkin, D. L. Sun, Z. W. Li, C. T. Lin, T. Wolf, B. Keimer, and A. V. Boris, Phys. Rev. B **88**, 184511 (2013).